\documentclass[10pt,twocolumn,twoside]{IEEEtran} 

\usepackage{multirow}
\usepackage[table,xcdraw]{xcolor}
\usepackage{graphicx}
\usepackage[bookmarks=false]{hyperref}
\usepackage{epstopdf}
\usepackage{booktabs}

\ifCLASSINFOpdf
\else
\fi
\ifCLASSOPTIONcompsoc
\usepackage[caption=false,font=normalsize,labelfont=sf,textfont=sf]{subfig}
\else
\usepackage[caption=false,font=footnotesize]{subfig}
\fi
\begin{document}
%
\title{Comments on ``Spatio-Temporal Gaussian Process Models for Extended and Group Object Tracking with Irregular Shapes"}
%
%
%

\author{Murat Kumru
	and Emre \"{O}zkan,~\IEEEmembership{Member,~IEEE}
\thanks{The authors are with the Department
	of Electrical and Electronics Engineering, Middle East Technical University, 06531, Ankara,
	Turkey (e-mail: kumru@metu.edu.tr; emreo@metu.edu.tr).}
}

%
%

\markboth{}%
{}
%



\maketitle

\begin{abstract}
	In the study ``Spatio-temporal Gaussian process models for extended and group object tracking with irregular shapes" (IEEE Trans. Veh. Tech., vol. 68, no. 3, pp. 2137--2151, Mar. 2019), the extended object tracking problem was tackled by an approach based on spatio-temporal Gaussian processes (STGP). 
	The performance of the proposed STGP-based trackers was comparatively evaluated through simulations and real data together with another state-of-the-art method (referred to as GP-EKF) proposed in ``Extended target tracking using Gaussian processes" (IEEE Trans. Signal Process., vol. 63, no. 16, pp. 4165--4178, Aug. 2015).  
	Unfortunately, we recognized that there are major errors in the implementation of the experiments presented in the STGP paper, which led to incorrect performance evaluation results.  
    In this correspondence, our aim is to share the correct results of these experiments and to respond to some claims regarding GP-EKF, which we believe, would contribute to a better understanding of the methods. 
\end{abstract}


%
\IEEEpeerreviewmaketitle

\section{Introduction}
In \cite{aftab2019spatio}, Aftab \textit{et al.} proposed a method to address the problem of tracking extended and group objects. 
The approach essentially relies on modeling the unknown extent by a spatio-temporal Gaussian process (STGP). 
The STGP description of the extent is approximated by a state space model which is then concatenated by the object kinematics. 
The resulting state vector including both the representation of the extent and the kinematics is recursively estimated by an extended Kalman filter (EKF) and a fixed-lag Rauch-Tung-Streibel (RTS) smoother. 
More specifically, the paper develops two methods named as STGP-EKF and STGP-RTSS; 
while STGP-EKF describes the filter, STGP-RTSS corresponds to the scheme making use of the smoother. 

To demonstrate the added value of the STGP-based extent modeling in object tracking, \cite{aftab2019spatio} regards a reference algorithm, GP-EKF, which was initially proposed by Wahlstr\"{o}m and \"{O}zkan in \cite{wahlstrom2015extended}. 
The performance of the aforementioned methods is comparatively assessed through various experiments in \cite{aftab2019spatio}. 
However, during our exploration of the corresponding source code\footnote{\url{https://figshare.shef.ac.uk/articles/STGP_for_Extended_Group_Object_Tracking_MATLAB_code_/7560314}}, we regrettably noticed that there are critical errors in the implementation of GP-EKF, which consequently led to incorrect performance evaluation and comparison in \cite{aftab2019spatio}. 
In particular, these errors can be listed as follows. 
\begin{itemize}
	\item The orientation of the object is not considered as a separate variable and not included in the state vector. 
	Instead, the orientation information is deduced from the velocity of the object. 
	This results in a completely different model than the original GP-EKF suggested in \cite{wahlstrom2015extended}. 
			
	\item Even more critically, there are multiple coding errors in the implementation of the intended version of GP-EKF (e.g., the state is incorrectly updated due to several mistakes in the computation of the partial derivatives required in the measurement update of the filter) which eventually cause unreliable estimation of the kinematics and the shape.  
\end{itemize}

\begin{figure*}[t]
   \centering
   \subfloat[Full GP]{\includegraphics[trim= 10 0 13 10,clip, width=0.25\textwidth]{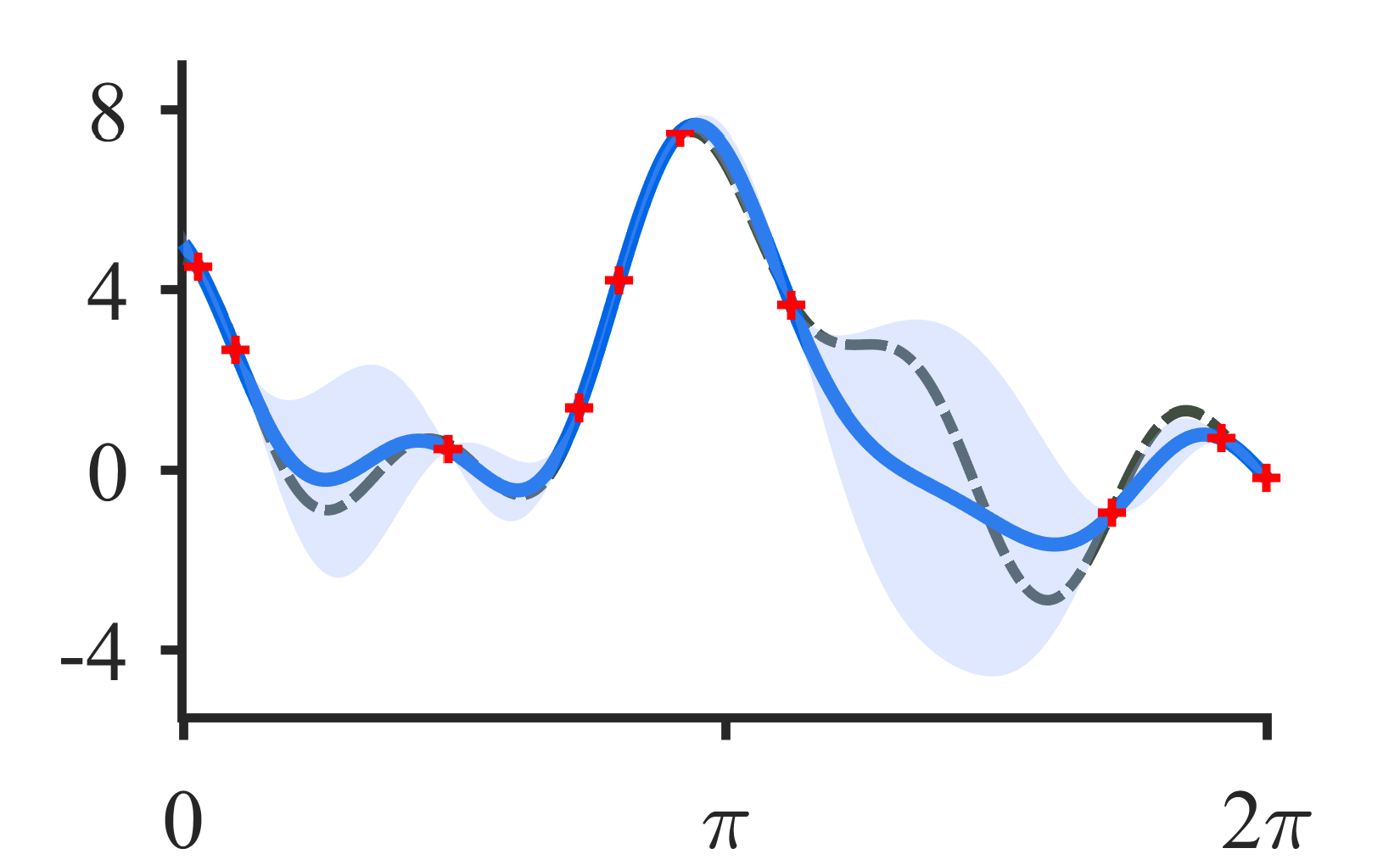}}
   \subfloat[RGP (5 basis points)]{\includegraphics[trim= 10 0 13 10,clip, width=0.25\textwidth]{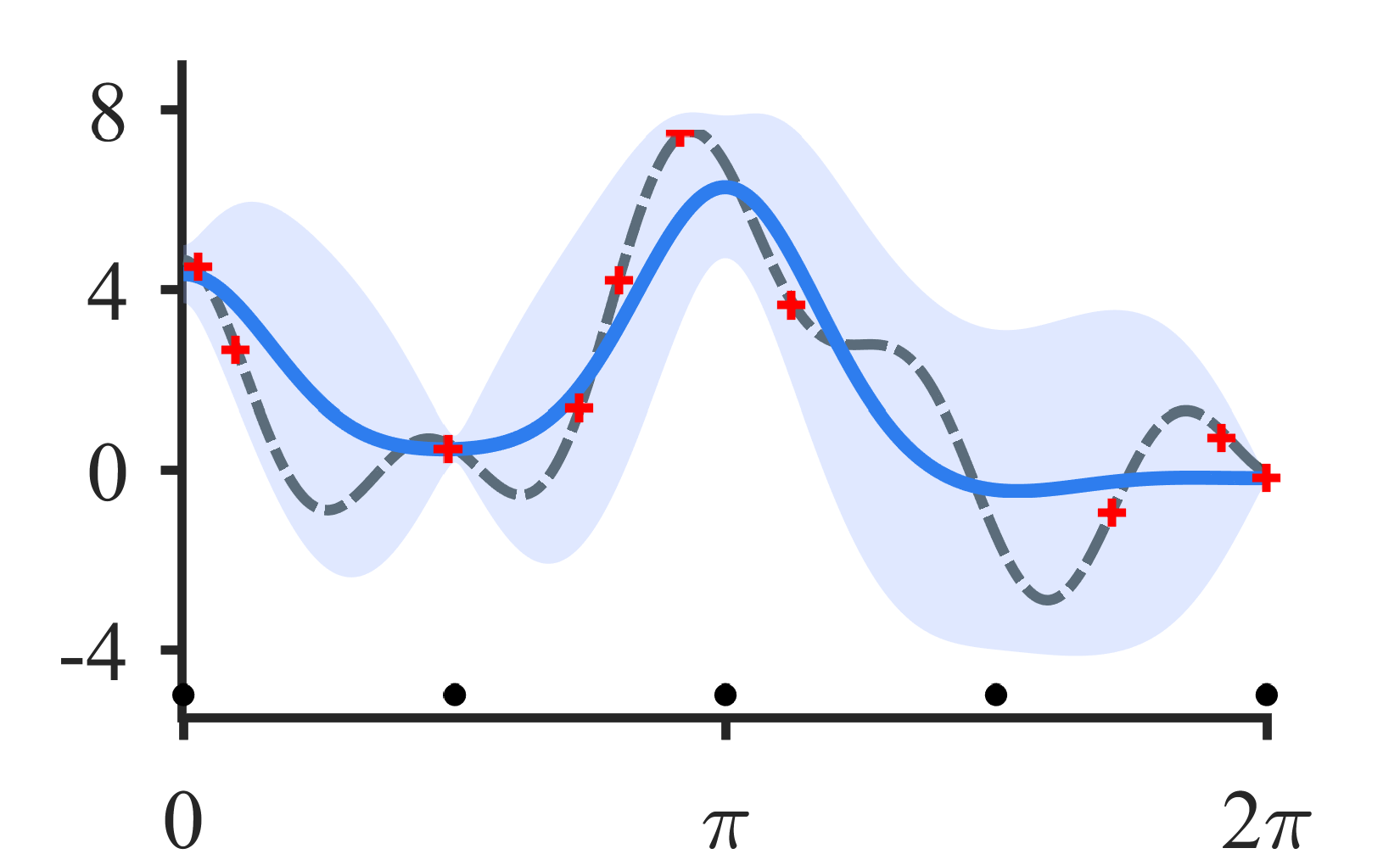}}
   \subfloat[RGP (10 basis points)]{\includegraphics[trim= 10 0 13 10,clip, width=0.25\textwidth]{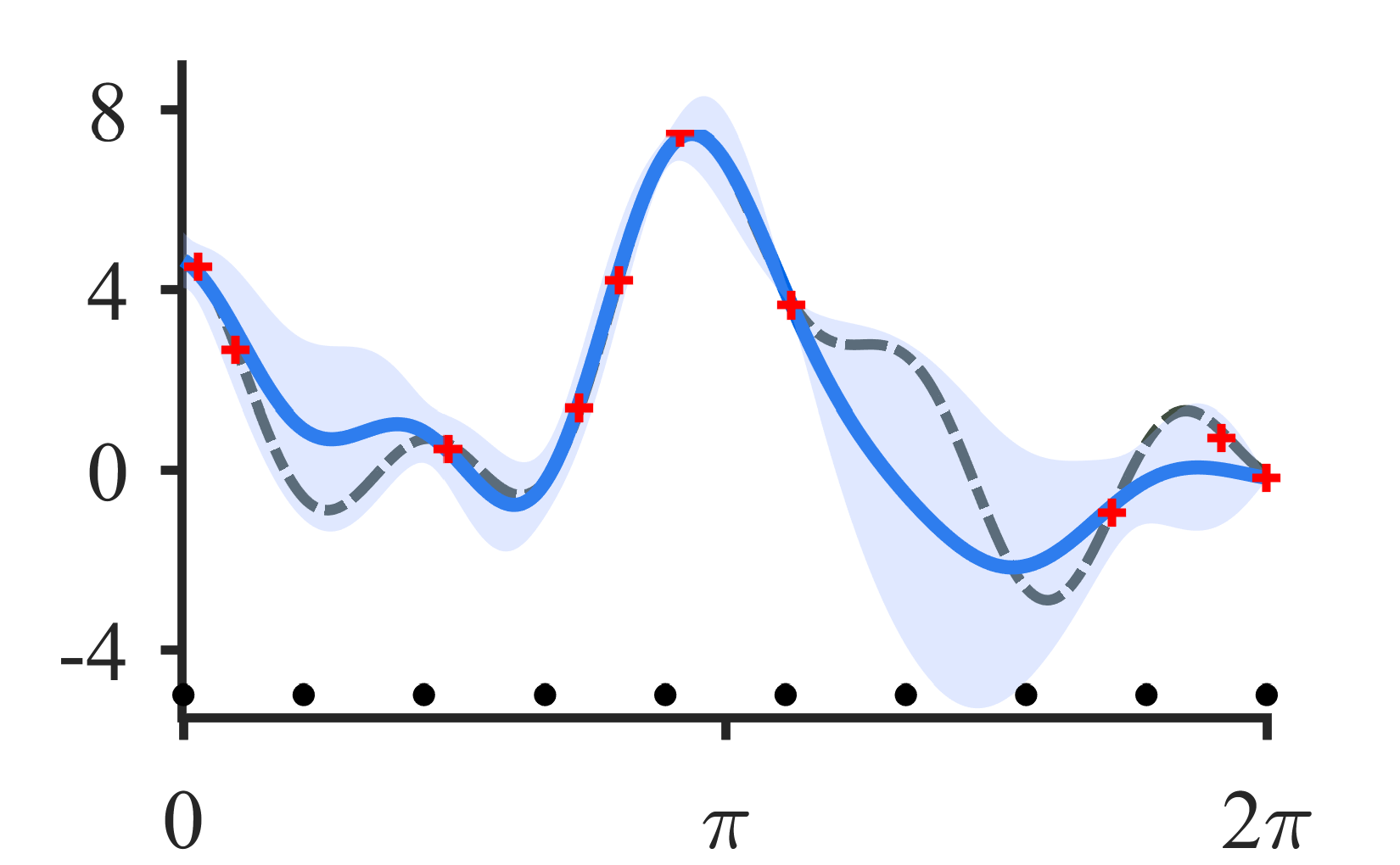}}
   \subfloat[RGP (20 basis points)]{\includegraphics[trim= 10 0 13 10,clip, width=0.25\textwidth]{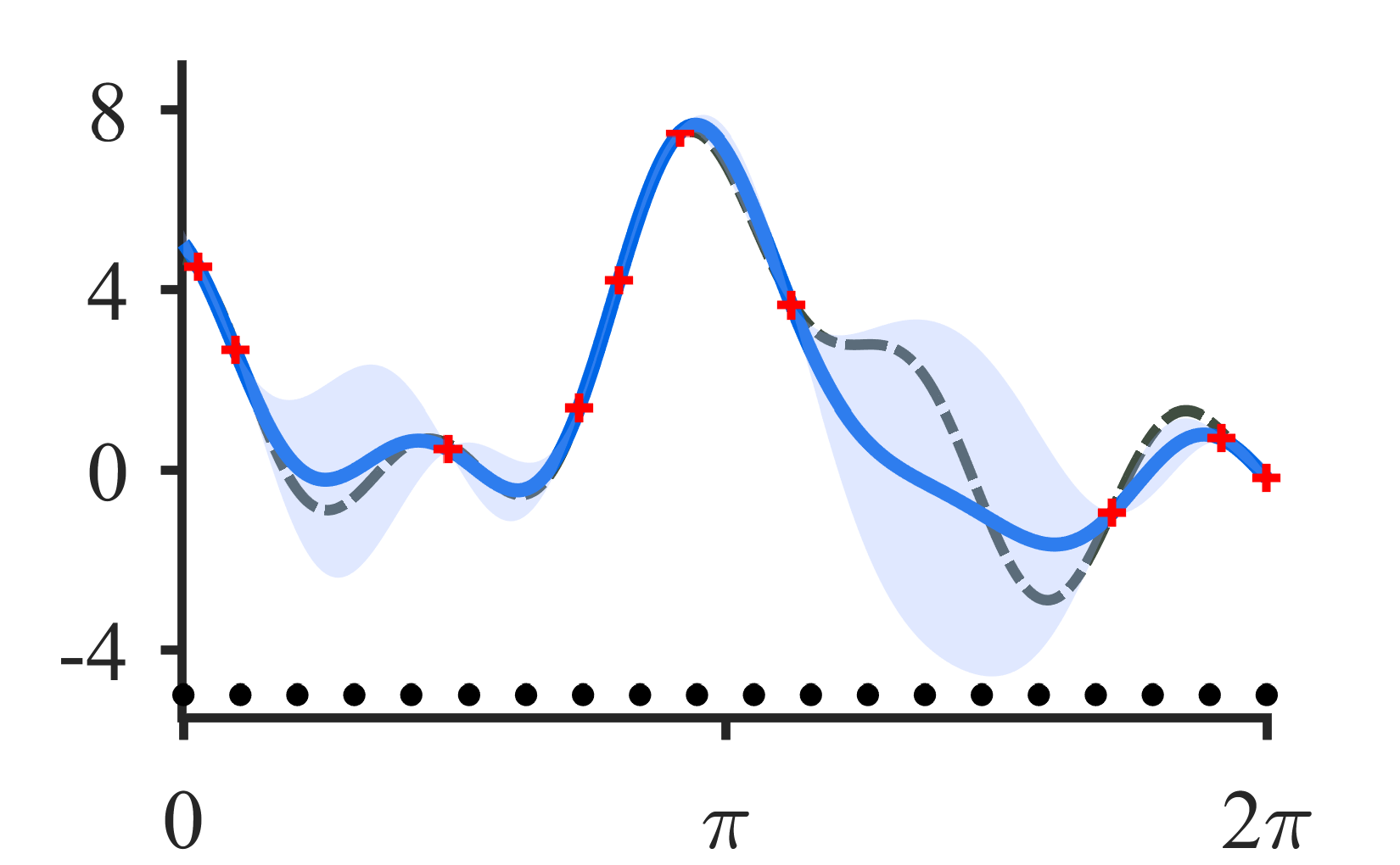}}
   \caption{Illustrative regression results of the full GP and three RGPs utilizing 5, 10 and 20 basis points. 
   The same set of measurements, which are indicated by red pluses, are provided to all methods. 
   The dashed line stands for the latent function. 
   The regression result is plot in solid blue with confidence region of 1 standard deviation. 
   In (b)-(d), the black dots over the \textit{x}-axis represent the locations of the basis points used by the corresponding RGP method.}
   \label{fig:GPRegression}
\end{figure*}

\begin{figure}[t]
   \centering
   \includegraphics[trim= 25 0 25 10,clip, width=0.95\columnwidth]{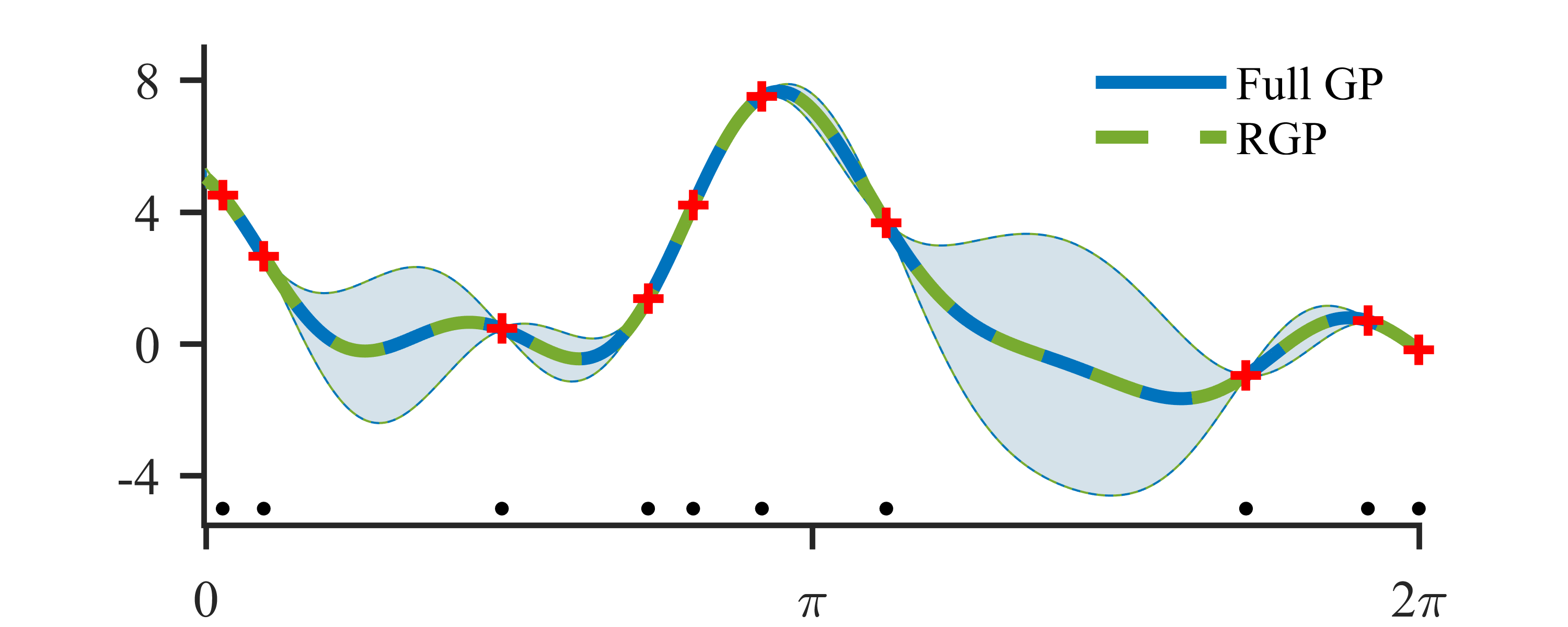}
   \caption{Regression results of the full GP and a specific RGP whose basis points are located at the inputs of the measurements as stated in \cite{wahlstrom2015extended}. 
   (Measurements are indicated by red pluses; the black dots over the \textit{x}-axis represent the locations of the basis points.)}
   \label{fig:GPRegressionComp}
\end{figure}

The errors directly affects the performance comparison results. Furthermore, we would like to reply to the following claims about GP-EKF asserted in \cite{aftab2019spatio}. 
\begin{itemize}
	\item On page 2, right column, in \cite{aftab2019spatio}, it is argued that ``Additionally, in [Wahlstr\"{o}m, \"{O}zkan] the GP based approach has been proposed equivalent to a batch GP regression without giving the theoretical explanation and the necessary conditions for the equivalence."
	
	\item[-] 
	This statement is incorrect as there is no discussion or implication about the equivalence of the proposed approach to a batch GP regression in \cite{wahlstrom2015extended}. 
    The only equivalence condition implied in the paper involves the approximation in the likelihood given in equation (13) of the paper, which reads $p\left(z_{k} | \mathbf{f}, z_{1: k-1}\right) \approx p\left(z_{k} | \mathbf{f}\right)$. 
    \cite{wahlstrom2015extended} further explains that 
    ``This approximation would be exact if the input values for $z_{1: k-1}$ were a subset of the input values for $\mathbf{f}$,  and it would be a good approximation if the input values for $z_{1: k-1}$ were close to those of $\mathbf{f}$ relative to the characteristic length scale of the covariance function."
    The remaining part of the derivations (regarding the GP model) does not involve any further approximations, and they follow from Bayes rule exactly. 
    
    To provide some insight into the consequences of the specified approximation, we illustrate a regression example in Figs. \ref{fig:GPRegression} and \ref{fig:GPRegressionComp}. 
    In the example, measurements are sequentially collected from a static function. 
    In Fig. \ref{fig:GPRegression}(a), the output of full GP regression is shown; for this case, batch processing of all measurements are performed to compute the posterior distribution. 
    Figs. \ref{fig:GPRegression}(b)-(d) exhibit the posterior distributions, which are recursively computed by the method proposed in \cite{wahlstrom2015extended} (it is denoted by RGP). 
    Note that the method utilizes a Kalman filter that regards a state vector consisting of the latent function values at the basis points. 
    Fig. \ref{fig:GPRegression} demonstrates that with the increasing number of basis points, the input of each measurement potentially gets closer to the basis points. 
    Hence, the resulting RGP approximates the full GP regression better. 
    Furthermore, in Fig. \ref{fig:GPRegressionComp}, the regression results of the full GP and a specific RGP are presented for the same example.
    In this case, the locations of the basis points of the RGP are selected to match the input points of the measurements. Even though the number of basis points is less than that of  Fig. \ref{fig:GPRegression}(d), the two regression results are shown to be identical as the condition that is stated in \cite{wahlstrom2015extended} is satisfied.

	
	\item
    On page 8, left column, in \cite{aftab2019spatio}, it is stated that ``A real-time recursive filter equivalent to a full GP regression has also been proposed in [Wahlstr\"{o}m, \"{O}zkan], [Hirscher \textit{et al.}]. 
    The mathematical equivalence of a full GP regression is a smoother rather than a filter [S\"{a}rkk\"{a}, Solin, Hartikainen]." 

    \item[-] 
    These arguments may be misleading in the sense that they necessitate a smoother for a full GP regression in all cases. 
    However, the method in \cite{wahlstrom2015extended} is suggested for online target tracking. 
    Target tracking mostly involves online applications that require the computation of the filtering density. 
    In an online application, where the objective is to obtain the filtering density $p(x_t|y_{1:t})$, no smoother is required for exact inference. 
    The filter and the smoother outputs are the same for estimating the state vector at time $t$ by using the measurements up to and including time $t$. 
    In that respect, online extent estimation at the last time instant is accomplished by using a filter, not a smoother. 
    This also applies to the STGP framework where the extent estimate at the last time instant is equal to the output of a filter which does not require a smoothing operation. 
\end{itemize}
In our opinion, both methods proposed in \cite{aftab2019spatio} and \cite{wahlstrom2015extended} have merits of their own. 
Using STGP modeling in extended object tracking is a smart idea and has advantages such as imposing time correlation of dynamic extents via GPs. 
However, in this commentary, we feel obliged to correct the previous findings of \cite{aftab2019spatio} and demonstrate that with the proper implementation of the method proposed in \cite{wahlstrom2015extended}, the claimed performance gain of the STGP-based tracker does not exist. 

\section{Results with Correct GP-EKF Implementation} \label{sec:Results}
In this section, we aim to present an accurate performance analysis of the aforementioned algorithms. 
To this end, we repeat the same experiments in \cite{aftab2019spatio} with the correct implementation\footnote{\url{https://github.com/Kumru/GPETT2D}} of GP-EKF.
To assess the performance of the methods, we utilize the same measures as defined in \cite{aftab2019spatio}, namely mean shape precision (P) and mean shape recall (R) and root-mean square error (RMSE) of the position and velocity of the center of the object (CoO). 
Notice that the considered algorithms do not directly estimate the CoO. 
Instead, the algorithms estimate the same physical quantity, which is called Internal Reference Point (IRP) in \cite{aftab2019spatio}, and target position in \cite{wahlstrom2015extended}. 
Consequently, for both methods, these position estimates are transformed into the CoO estimates by using the equations given in Section IV-G in \cite{aftab2019spatio}.

\subsection{Simulation Experiments}
In this subsection, we present the results obtained from simulation experiments. 
The setup of the simulations and the parameters of the STGP-based trackers are exactly the same with those in \cite{aftab2019spatio}. 
More specifically, we have directly used the simulation environment and the implementation of the STGP-based trackers provided by the source code of the reference paper. 
All simulation experiments are repeated 100 times and the presented numbers are obtained by averaging these Monte Carlo (MC) runs. 

For GP-EKF, the process noise standard deviations for the position and the orientation angle are used as $\sigma_q=1$, ${\sigma_{q^\psi}=0.0001}$, respectively; 
the hyperparameters of the GP model are tuned to ${\sigma_f=2}$, ${\sigma_r=0.8}$ and ${l=\pi/10}$; 
${\alpha=0.004}$ is used as the forgetting factor in the extent dynamics. 

\begin{figure*}[h]
	\centering
	\includegraphics[width=0.7\textwidth] {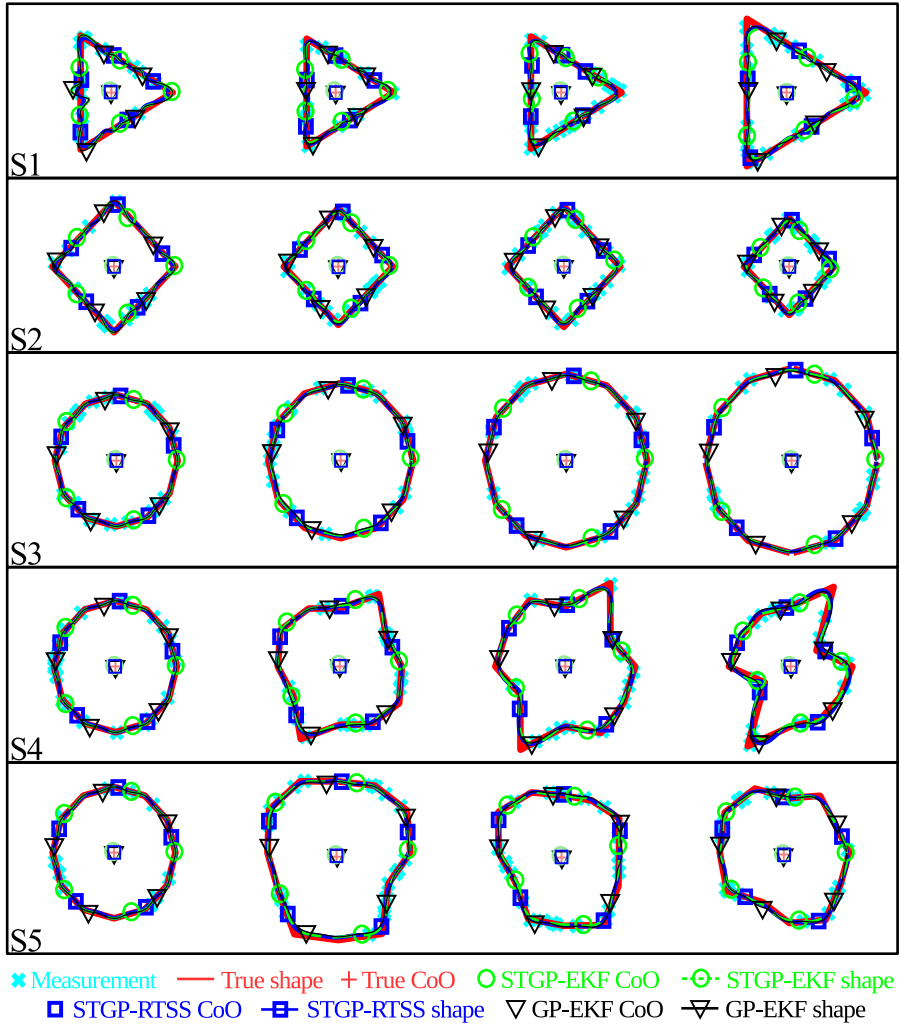}
	\caption{Example snapshots taken at frames $\{1,50,150,230\}$ during a typical simulation experiment. The figure basically reproduces its equivalent, \cite[Fig.~6]{aftab2019spatio}, by relying on the correct implementation of GP-EKF. }
	\centering
	\label{fig:snapshots}
\end{figure*}

Table \ref{table:rmse} presents the measures evaluated for the STGP-based trackers. 
To be consistent with \cite{aftab2019spatio}, Table \ref{table:mpi} reveals the mean percentage improvement of the STGP-based trackers over GP-EKF. 
Following the same notation with \cite{aftab2019spatio}, a positive value indicates that the STGP-based tracker performs better than GP-EKF; on the contrary, a negative value implies that GP-EKF outperforms the STGP-based tracker. 
In addition, Fig. \ref{fig:snapshots} exhibits the typical outputs of the algorithms acquired at the specific frames of the experiments. 

The numbers displayed in Table \ref{table:rmse} are expectedly consistent with those given in \cite[Table~II]{aftab2019spatio} as we have not modified the implementation of the STGP-based trackers. 
On the other hand, the following points should be highlighted as a result of the comparison between Table \ref{table:mpi}, Fig. \ref{fig:snapshots} and their respective counterparts in \cite{aftab2019spatio}. 
\begin{itemize}
	\item In \cite{aftab2019spatio}, it is claimed that STGP is able to model the evolution of the extent more precisely than the other methods so that the proposed STGP-based tracking algorithms can achieve enhanced shape estimation performance. 
	The erroneous results obtained by the incorrect implementation of GP-EKF support these arguments. 
	For example, \cite{aftab2019spatio} reports the shape estimates of GP-EKF to be less accurate relying on the inconsistencies in \cite[Fig.~6]{aftab2019spatio}. 
	Similarly, \cite[Table~III]{aftab2019spatio} reveals that the STGP-based trackers exhibit favorable performance in the precision and recall measures. 
	However, the precision and recall measures in Table \ref{table:mpi} indicate that the STGP-based trackers do not provide any relative gain in shape estimation. 
	This is also visualized by Fig. \ref{fig:snapshots} which demonstrates that GP-EKF is able to successfully estimate the shape of the object for all scenarios. 
			
	\item Table \ref{table:mpi} suggests that GP-EKF generates substantially enhanced position estimates compared to STGP-EKF for all shape models. 
	Specifically, GP-EKF achieves up to \%390 improvement in the position accuracy. 
	In contrast, \cite[Table~III]{aftab2019spatio} incorrectly reports that STGP-EKF is superior to GP-EKF in position estimation by up to \%85. 	 	 	
		
	\item Table \ref{table:mpi} shows that STGP-RTSS generally outperforms GP-EKF in the estimation of the kinematics. 
	However, the level of improvement obtained by STGP-RTSS is observed to be far below than that specified by \cite[Table~III]{aftab2019spatio}. 		
	Moreover, it is important to bear in mind that STGP-RTSS deploys a fixed-lag RTS smoother in its architecture. 
	The smoother regards a unified state space model consisting of both the extent representation and the kinematics. 
	Therefore, the kinematic estimates produced by STGP-RTSS benefit from the information embedded in future measurements which is not provided to GP-EKF. 
	Consequently, the improved performance of STGP-RTSS in kinematic estimation can also stem from this additional information, and it is not suited to prove the added value of STGP modeling in object tracking.
    In other words, it is simply not appropriate to compare a smoother to a filter; instead, a reasonable comparison can be made between two smoothers. 
    In this regard, we implemented a smoother denoted as GP-RTSS in order to make a fairer comparison and to better understand the performance difference of the methods. 
    Similar to STGP-RTSS, GP-RTSS is basically realized by a fixed-lag RTS smoother; however, it relies on the GP representation of the object extent suggested in \cite{wahlstrom2015extended}. 
    The performance of the method is comparatively evaluated through the aforementioned simulation experiments. 
    Note that the lag value of GP-RTSS is set to be equal to that of STGP-RTSS, and the parameters of GP-RTSS are identical to those of GP-EKF. 
    Table \ref{table:mpi2} demonstrates the mean percentage improvement of the STGP-based trackers over GP-RTSS. 
    It is observed that GP-RTSS significantly outperforms both of the STGP-based trackers in the accuracy of the kinematic estimates for all scenarios. 
    In particular, the comparison between GP-RTSS and STGP-RTSS shows that once the methods are provided with the same amount of information, the one relying on the GP model is superior to the other in kinematic estimation. 
    
    \item Lastly, it is worth mentioning that the STGP-based trackers proposed in \cite{aftab2019spatio} do not estimate the orientation of the object as they explicitly formulate the model for fixed orientation. 
    Consequently, they are not suitable for the scenarios that involves object rotation without modifications. 
	Note that the experiments conducted in \cite{aftab2019spatio} (and repeated in this study) examine only objects with fixed orientation and hence do not pose a problem for STGP-EKF and STGP-RTSS. 
	On the other hand, GP-EKF and GP-RTSS estimate the orientation together with the kinematics and the shape of the object. 
	While this leads to a more flexible scheme that can generalize to a broader class of scenarios, it also introduces additional uncertainty due to the unknown orientation. 
	That being said, throughout the experiments, it is shown that the GP-based trackers are able to estimate the unknown orientation successfully and they provide a preferable performance while trying to solve a problem with higher uncertainty.	
\end{itemize}

\begin{table}[ht]
		\centering	
		\caption{Performance Measures Evaluated for the STGP-based Trackers (Simulations)}
		\label{table:rmse}
		\begin{tabular}{|c|c|l|l|l|l|l|l|}
			\hline
			\multicolumn{2}{|c|}{} & \multicolumn{1}{c|}{} & \multicolumn{5}{c|}{\textbf{Shape Model}} \\ \cline{4-8} 
			\multicolumn{2}{|c|}{\multirow{-2}{*}{\textbf{Measure}}} & \multicolumn{1}{c|}{\multirow{-2}{*}{\textbf{Method}}} & \multicolumn{1}{c|}{\textbf{S1}} & \multicolumn{1}{c|}{\textbf{S2}} & \multicolumn{1}{c|}{\textbf{S3}} & \multicolumn{1}{c|}{\textbf{S4}} & \multicolumn{1}{c|}{\textbf{S5}} \\ \hline
			&  & STGP-EKF & \cellcolor[HTML]{FFFFFF}0.27 & 0.13 & 0.08 & 0.12 & 0.13 \\
			& \multirow{-2}{*}{\begin{tabular}[c]{@{}c@{}}$x$\\ (m)\end{tabular}} & \cellcolor[HTML]{C0C0C0}STGP-RTSS & \cellcolor[HTML]{C0C0C0}0.15 & \cellcolor[HTML]{C0C0C0}0.08 & \cellcolor[HTML]{C0C0C0}0.05 & \cellcolor[HTML]{C0C0C0}0.06 & \cellcolor[HTML]{C0C0C0}0.07 \\ \cline{2-8} 
			&  & STGP-EKF & 0.49 & 0.36 & 0.08 & 0.13 & 0.14 \\
			& \multirow{-2}{*}{\begin{tabular}[c]{@{}c@{}}$y$\\ (m)\end{tabular}} & \cellcolor[HTML]{C0C0C0}STGP-RTSS & \cellcolor[HTML]{C0C0C0}0.32 & \cellcolor[HTML]{C0C0C0}0.29 & \cellcolor[HTML]{C0C0C0}0.04 & \cellcolor[HTML]{C0C0C0}0.07 & \cellcolor[HTML]{C0C0C0}0.08 \\ \cline{2-8} 
			&  & STGP-EKF & 1.11 & 0.84 & 0.74 & 0.86 & 0.89 \\
			& \multirow{-2}{*}{\begin{tabular}[c]{@{}c@{}}$\dot{x}$\\ (m/s)\end{tabular}} & \cellcolor[HTML]{C0C0C0}STGP-RTSS & \cellcolor[HTML]{C0C0C0}0.39 & \cellcolor[HTML]{C0C0C0}0.30 & \cellcolor[HTML]{C0C0C0}0.25 & \cellcolor[HTML]{C0C0C0}0.29 & \cellcolor[HTML]{C0C0C0}0.36 \\ \cline{2-8} 
			&  & STGP-EKF & 1.09 & 0.90 & 0.73 & 0.87 & 0.89 \\
			\multirow{-8}{*}{\rotatebox{90}{RMSE }} & \multirow{-2}{*}{\begin{tabular}[c]{@{}c@{}}$\dot{y}$\\ (m/s)\end{tabular}} & \cellcolor[HTML]{C0C0C0}STGP-RTSS & \cellcolor[HTML]{C0C0C0}0.39 & \cellcolor[HTML]{C0C0C0}0.43 & \cellcolor[HTML]{C0C0C0}0.25 & \cellcolor[HTML]{C0C0C0}0.28 & \cellcolor[HTML]{C0C0C0}0.36 \\ \hline
			\multicolumn{2}{|c|}{} & STGP-EKF & 0.98 & 0.99 & 1.00 & 0.99 & 0.99 \\
			\multicolumn{2}{|c|}{\multirow{-2}{*}{$P$}} & \cellcolor[HTML]{C0C0C0}STGP-RTSS & \cellcolor[HTML]{C0C0C0}0.97 & \cellcolor[HTML]{C0C0C0}0.99 & \cellcolor[HTML]{C0C0C0}0.99 & \cellcolor[HTML]{C0C0C0}0.99 & \cellcolor[HTML]{C0C0C0}0.99 \\ \hline
			\multicolumn{2}{|c|}{} & STGP-EKF & 0.96 & 0.98 & 0.99 & 0.98 & 0.98 \\
			\multicolumn{2}{|c|}{\multirow{-2}{*}{$R$}} & \cellcolor[HTML]{C0C0C0}STGP-RTSS & \cellcolor[HTML]{C0C0C0}0.98 & \cellcolor[HTML]{C0C0C0}0.99 & \cellcolor[HTML]{C0C0C0}0.99 & \cellcolor[HTML]{C0C0C0}0.99 & \cellcolor[HTML]{C0C0C0}0.99 \\ \hline
		\end{tabular}	
	\end{table}

\begin{table}[ht]
	\caption{Mean Percentage Improvements of the STGP-based Trackers over GP-EKF (Simulations)}
	\centering
	\label{table:mpi}
	\resizebox{\columnwidth}{!}{
		\begin{tabular}{|c|c|l|c|c|c|c|c|}
    \hline
    \multicolumn{2}{|c|}{{\color[HTML]{333333} }} & \multicolumn{1}{c|}{{\color[HTML]{333333} }} & \multicolumn{5}{c|}{{\color[HTML]{333333} \textbf{Shape Model}}} \\ \cline{4-8} 
    \multicolumn{2}{|c|}{\multirow{-2}{*}{{\color[HTML]{333333} \textbf{Measure}}}} & \multicolumn{1}{c|}{\multirow{-2}{*}{{\color[HTML]{333333} \textbf{Method}}}} & {\color[HTML]{333333} \textbf{S1}} & {\color[HTML]{333333} \textbf{S2}} & {\color[HTML]{333333} \textbf{S3}} & {\color[HTML]{333333} \textbf{S4}} & {\color[HTML]{333333} \textbf{S5}} \\ \hline
    {\color[HTML]{333333} } & {\color[HTML]{333333} } & {\color[HTML]{333333} STGP-EKF} & \cellcolor[HTML]{FFFFFF}{\color[HTML]{333333} \textbf{-167.33}} & {\color[HTML]{333333} \textbf{-77.73}} & {\color[HTML]{333333} \textbf{-33.87}} & {\color[HTML]{333333} \textbf{-56.72}} & {\color[HTML]{333333} \textbf{-41.92}} \\
    {\color[HTML]{333333} } & \multirow{-2}{*}{{\color[HTML]{333333} \begin{tabular}[c]{@{}c@{}}$x$\end{tabular}}} & \cellcolor[HTML]{C0C0C0}{\color[HTML]{333333} STGP-RTSS} & \cellcolor[HTML]{C0C0C0}{\color[HTML]{333333} \textbf{-50.86}} & \cellcolor[HTML]{C0C0C0}{\color[HTML]{333333} \textbf{-13.85}} & \cellcolor[HTML]{C0C0C0}{\color[HTML]{333333} 28.43} & \cellcolor[HTML]{C0C0C0}{\color[HTML]{333333} 24.41} & \cellcolor[HTML]{C0C0C0}{\color[HTML]{333333} 18.32} \\ \cline{2-8} 
    {\color[HTML]{333333} } & {\color[HTML]{333333} } & {\color[HTML]{333333} STGP-EKF} & {\color[HTML]{333333} \textbf{-394.22}} & {\color[HTML]{333333} \textbf{-381.25}} & {\color[HTML]{333333} \textbf{-25.82}} & {\color[HTML]{333333} \textbf{-66.50}} & {\color[HTML]{333333} \textbf{-68.34}} \\
    {\color[HTML]{333333} } & \multirow{-2}{*}{{\color[HTML]{333333} \begin{tabular}[c]{@{}c@{}}$y$\end{tabular}}} & \cellcolor[HTML]{C0C0C0}{\color[HTML]{333333} STGP-RTSS} & \cellcolor[HTML]{C0C0C0}{\color[HTML]{333333} \textbf{-211.86}} & \cellcolor[HTML]{C0C0C0}{\color[HTML]{333333} \textbf{-292.67}} & \cellcolor[HTML]{C0C0C0}{\color[HTML]{333333} 36.09} & \cellcolor[HTML]{C0C0C0}{\color[HTML]{333333} 17.60} & \cellcolor[HTML]{C0C0C0}{\color[HTML]{333333} 9.04} \\ \cline{2-8} 
    {\color[HTML]{333333} } & {\color[HTML]{333333} } & {\color[HTML]{333333} STGP-EKF} & {\color[HTML]{333333} 53.64} & {\color[HTML]{333333} 49.79} & {\color[HTML]{333333} 40.01} & {\color[HTML]{333333} 48.04} & {\color[HTML]{333333} 44.04} \\
    {\color[HTML]{333333} } & \multirow{-2}{*}{{\color[HTML]{333333} \begin{tabular}[c]{@{}c@{}}$\dot{x}$\end{tabular}}} & \cellcolor[HTML]{C0C0C0}{\color[HTML]{333333} STGP-RTSS} & \cellcolor[HTML]{C0C0C0}{\color[HTML]{333333} 82.98} & \cellcolor[HTML]{C0C0C0}{\color[HTML]{333333} 82.22} & \cellcolor[HTML]{C0C0C0}{\color[HTML]{333333} 79.46} & \cellcolor[HTML]{C0C0C0}{\color[HTML]{333333} 82.37} & \cellcolor[HTML]{C0C0C0}{\color[HTML]{333333} 77.61} \\ \cline{2-8} 
    {\color[HTML]{333333} } & {\color[HTML]{333333} } & {\color[HTML]{333333} STGP-EKF} & {\color[HTML]{333333} 53.69} & {\color[HTML]{333333} 45.55} & {\color[HTML]{333333} 40.95} & {\color[HTML]{333333} 47.22} & {\color[HTML]{333333} 43.16} \\
    \multirow{-8}{*}{{\color[HTML]{333333} \rotatebox{90}{RMSE }}} & \multirow{-2}{*}{{\color[HTML]{333333} \begin{tabular}[c]{@{}c@{}}$\dot{y}$\end{tabular}}} & \cellcolor[HTML]{C0C0C0}{\color[HTML]{333333} STGP-RTSS} & \cellcolor[HTML]{C0C0C0}{\color[HTML]{333333} 83.60} & \cellcolor[HTML]{C0C0C0}{\color[HTML]{333333} 73.74} & \cellcolor[HTML]{C0C0C0}{\color[HTML]{333333} 79.51} & \cellcolor[HTML]{C0C0C0}{\color[HTML]{333333} 83.03} & \cellcolor[HTML]{C0C0C0}{\color[HTML]{333333} 77.23} \\ \hline
    \multicolumn{2}{|c|}{{\color[HTML]{333333} }} & {\color[HTML]{333333} STGP-EKF} & {\color[HTML]{333333} \textbf{-0.04}} & {\color[HTML]{333333} 0.50} & {\color[HTML]{333333} 0.46} & {\color[HTML]{333333} 0.45} & {\color[HTML]{333333} 0.43} \\
    \multicolumn{2}{|c|}{\multirow{-2}{*}{{\color[HTML]{333333} $P$}}} & \cellcolor[HTML]{C0C0C0}{\color[HTML]{333333} STGP-RTSS} & \cellcolor[HTML]{C0C0C0}{\color[HTML]{333333} \textbf{-0.65}} & \cellcolor[HTML]{C0C0C0}{\color[HTML]{333333} \textbf{-0.08}} & \cellcolor[HTML]{C0C0C0}{\color[HTML]{333333} 0.14} & \cellcolor[HTML]{C0C0C0}{\color[HTML]{333333} 0.06} & \cellcolor[HTML]{C0C0C0}{\color[HTML]{333333} 0.03} \\ \hline
    \multicolumn{2}{|c|}{{\color[HTML]{333333} }} & {\color[HTML]{333333} STGP-EKF} & {\color[HTML]{333333} \textbf{-2.23}} & {\color[HTML]{333333} \textbf{-1.28}} & {\color[HTML]{333333} \textbf{-0.82}} & {\color[HTML]{333333} \textbf{-1.00}} & {\color[HTML]{333333} \textbf{-0.96}} \\
    \multicolumn{2}{|c|}{\multirow{-2}{*}{{\color[HTML]{333333} $R$}}} & \cellcolor[HTML]{C0C0C0}{\color[HTML]{333333} STGP-RTSS} & \cellcolor[HTML]{C0C0C0}{\color[HTML]{333333} \textbf{-0.42}} & \cellcolor[HTML]{C0C0C0}{\color[HTML]{333333} \textbf{-0.11}} & \cellcolor[HTML]{C0C0C0}{\color[HTML]{333333} 0.04} & \cellcolor[HTML]{C0C0C0}{\color[HTML]{333333} 0.06} & \cellcolor[HTML]{C0C0C0}{\color[HTML]{333333} 0.06} \\ \hline
    \end{tabular}}
\end{table}

\begin{table}[ht]
	\caption{Mean Percentage Improvements of the STGP-based Trackers over GP-RTSS (Simulations)}
	\centering
	\label{table:mpi2}
	\resizebox{\columnwidth}{!}{
		\begin{tabular}{|c|c|l|c|c|c|c|c|}
        \hline
        \multicolumn{2}{|c|}{{\color[HTML]{333333} }} & \multicolumn{1}{c|}{{\color[HTML]{333333} }} & \multicolumn{5}{c|}{{\color[HTML]{333333} \textbf{Shape Model}}} \\ \cline{4-8} 
        \multicolumn{2}{|c|}{\multirow{-2}{*}{{\color[HTML]{333333} \textbf{Measure}}}} & \multicolumn{1}{c|}{\multirow{-2}{*}{{\color[HTML]{333333} \textbf{Method}}}} & {\color[HTML]{333333} \textbf{S1}} & {\color[HTML]{333333} \textbf{S2}} & {\color[HTML]{333333} \textbf{S3}} & {\color[HTML]{333333} \textbf{S4}} & {\color[HTML]{333333} \textbf{S5}} \\ \hline
        {\color[HTML]{333333} } & {\color[HTML]{333333} } & {\color[HTML]{333333} STGP-EKF} & \cellcolor[HTML]{FFFFFF}{\color[HTML]{333333} \textbf{-364.30}} & {\color[HTML]{333333} \textbf{-241.74}} & {\color[HTML]{333333} \textbf{-193.50}} & {\color[HTML]{333333} \textbf{-225.84}} & {\color[HTML]{333333} \textbf{-120.36}} \\
        {\color[HTML]{333333} } & \multirow{-2}{*}{{\color[HTML]{333333} $x$}} & \cellcolor[HTML]{C0C0C0}{\color[HTML]{333333} STGP-RTSS} & \cellcolor[HTML]{C0C0C0}{\color[HTML]{333333} \textbf{-162.01}} & \cellcolor[HTML]{C0C0C0}{\color[HTML]{333333} \textbf{-118.91}} & \cellcolor[HTML]{C0C0C0}{\color[HTML]{333333} \textbf{-56.92}} & \cellcolor[HTML]{C0C0C0}{\color[HTML]{333333} \textbf{-57.16}} & \cellcolor[HTML]{C0C0C0}{\color[HTML]{333333} \textbf{-26.83}} \\ \cline{2-8} 
        {\color[HTML]{333333} } & {\color[HTML]{333333} } & {\color[HTML]{333333} STGP-EKF} & {\color[HTML]{333333} \textbf{-763.16}} & {\color[HTML]{333333} \textbf{-815.91}} & {\color[HTML]{333333} \textbf{-184.55}} & {\color[HTML]{333333} \textbf{-249.27}} & {\color[HTML]{333333} \textbf{-173.95}} \\
        {\color[HTML]{333333} } & \multirow{-2}{*}{{\color[HTML]{333333} $y$}} & \cellcolor[HTML]{C0C0C0}{\color[HTML]{333333} STGP-RTSS} & \cellcolor[HTML]{C0C0C0}{\color[HTML]{333333} \textbf{-444.67}} & \cellcolor[HTML]{C0C0C0}{\color[HTML]{333333} \textbf{-647.32}} & \cellcolor[HTML]{C0C0C0}{\color[HTML]{333333} \textbf{-44.54}} & \cellcolor[HTML]{C0C0C0}{\color[HTML]{333333} \textbf{-72.85}} & \cellcolor[HTML]{C0C0C0}{\color[HTML]{333333} \textbf{-48.03}} \\ \cline{2-8} 
        {\color[HTML]{333333} } & {\color[HTML]{333333} } & {\color[HTML]{333333} STGP-EKF} & {\color[HTML]{333333} \textbf{-100.87}} & {\color[HTML]{333333} \textbf{-148.17}} & {\color[HTML]{333333} \textbf{-212.99}} & {\color[HTML]{333333} \textbf{-176.51}} & {\color[HTML]{333333} \textbf{-163.26}} \\
        {\color[HTML]{333333} } & \multirow{-2}{*}{{\color[HTML]{333333} $\dot{x}$}} & \cellcolor[HTML]{C0C0C0}{\color[HTML]{333333} STGP-RTSS} & \cellcolor[HTML]{C0C0C0}{\color[HTML]{333333} 26.25} & \cellcolor[HTML]{C0C0C0}{\color[HTML]{333333} 12.13} & \cellcolor[HTML]{C0C0C0}{\color[HTML]{333333} \textbf{-7.17}} & \cellcolor[HTML]{C0C0C0}{\color[HTML]{333333} 6.16} & \cellcolor[HTML]{C0C0C0}{\color[HTML]{333333} \textbf{-5.32}} \\ \cline{2-8} 
        {\color[HTML]{333333} } & {\color[HTML]{333333} } & {\color[HTML]{333333} STGP-EKF} & {\color[HTML]{333333} \textbf{-102.16}} & {\color[HTML]{333333} \textbf{-161.41}} & {\color[HTML]{333333} \textbf{-198.05}} & {\color[HTML]{333333} \textbf{-180.13}} & {\color[HTML]{333333} \textbf{-164.98}} \\
        \multirow{-8}{*}{{\color[HTML]{333333} \rotatebox{90}{RMSE }}} & \multirow{-2}{*}{{\color[HTML]{333333} $\dot{y}$}} & \cellcolor[HTML]{C0C0C0}{\color[HTML]{333333} STGP-RTSS} & \cellcolor[HTML]{C0C0C0}{\color[HTML]{333333} 28.42} & \cellcolor[HTML]{C0C0C0}{\color[HTML]{333333} \textbf{-26.06}} & \cellcolor[HTML]{C0C0C0}{\color[HTML]{333333} \textbf{-3.41}} & \cellcolor[HTML]{C0C0C0}{\color[HTML]{333333} 9.93} & \cellcolor[HTML]{C0C0C0}{\color[HTML]{333333} \textbf{-6.15}} \\ \hline
        \multicolumn{2}{|c|}{{\color[HTML]{333333} }} & {\color[HTML]{333333} STGP-EKF} & {\color[HTML]{333333} \textbf{-0.23}} & {\color[HTML]{333333} 0.25} & {\color[HTML]{333333} 0.13} & {\color[HTML]{333333} 0.18} & {\color[HTML]{333333} 0.17} \\
        \multicolumn{2}{|c|}{\multirow{-2}{*}{{\color[HTML]{333333} $P$}}} & \cellcolor[HTML]{C0C0C0}{\color[HTML]{333333} STGP-RTSS} & \cellcolor[HTML]{C0C0C0}{\color[HTML]{333333} \textbf{-0.84}} & \cellcolor[HTML]{C0C0C0}{\color[HTML]{333333} \textbf{-0.33}} & \cellcolor[HTML]{C0C0C0}{\color[HTML]{333333} \textbf{-0.19}} & \cellcolor[HTML]{C0C0C0}{\color[HTML]{333333} \textbf{-0.21}} & \cellcolor[HTML]{C0C0C0}{\color[HTML]{333333} \textbf{-0.22}} \\ \hline
        \multicolumn{2}{|c|}{{\color[HTML]{333333} }} & {\color[HTML]{333333} STGP-EKF} & {\color[HTML]{333333} \textbf{-2.43}} & {\color[HTML]{333333} \textbf{-1.46}} & {\color[HTML]{333333} \textbf{-1.10}} & {\color[HTML]{333333} \textbf{-1.30}} & {\color[HTML]{333333} \textbf{-1.18}} \\
        \multicolumn{2}{|c|}{\multirow{-2}{*}{{\color[HTML]{333333} $R$}}} & \cellcolor[HTML]{C0C0C0}{\color[HTML]{333333} STGP-RTSS} & \cellcolor[HTML]{C0C0C0}{\color[HTML]{333333} \textbf{-0.63}} & \cellcolor[HTML]{C0C0C0}{\color[HTML]{333333} \textbf{-0.29}} & \cellcolor[HTML]{C0C0C0}{\color[HTML]{333333} \textbf{-0.24}} & \cellcolor[HTML]{C0C0C0}{\color[HTML]{333333} \textbf{-0.24}} & \cellcolor[HTML]{C0C0C0}{\color[HTML]{333333} \textbf{-0.16}} \\ \hline
        \end{tabular}}
\end{table}

\subsection{Experiments with Real Measurements}
In this subsection, the performance of the algorithms is compared using real measurements. 
The considered experiments are identical to those given in Section IV-D in \cite{aftab2019spatio}. 
We have directly employed the source code of the reference paper for the construction of the experimental setup and the implementation the STGP-based trackers. 
For the GP-based trackers, the process noise standard deviations for the position and the orientation angle are used as $\sigma_q=3$, ${\sigma_{q^\psi}=0.000001}$, respectively; 
the hyperparameters of the GP model are tuned to ${\sigma_f=4}$, ${\sigma_r=0.8}$ and ${l=\pi/12}$; 
${\alpha=0.08}$ is used as the forgetting factor in the extent dynamics. 

\begin{figure}[h]
	\centering
	\includegraphics[trim=0 37 0 0,clip, width=1\columnwidth] {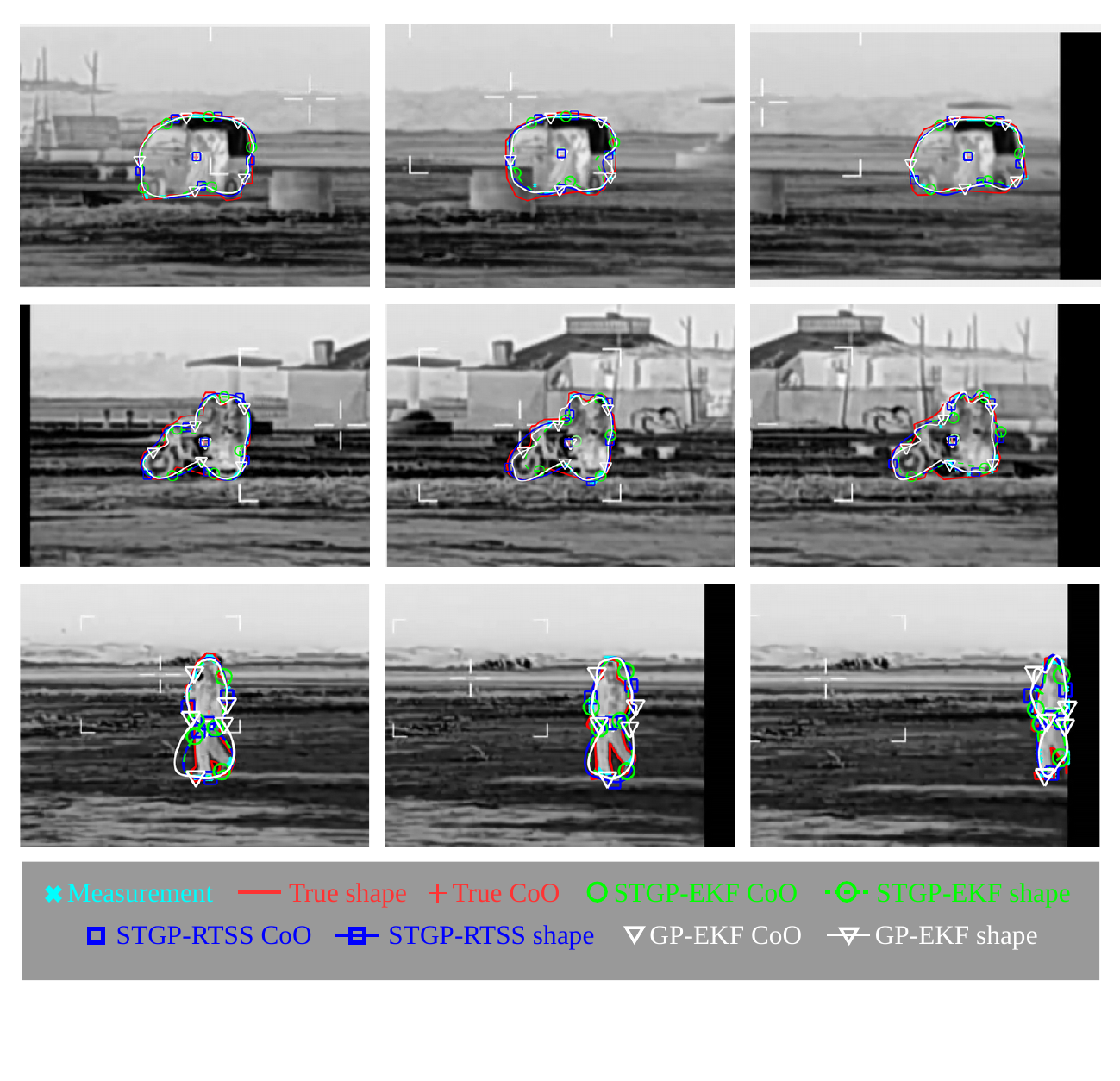}
	\caption{Snapshots of three frames during the experiments with real data (top: rickshaw, middle: motorcycle, bottom: pedestrian). The figure basically reproduces its equivalent, \cite[Fig.~11]{aftab2019spatio}, by relying on the correct implementation of GP-EKF. }
	\centering
	\label{fig:RealResults}
    \end{figure}

Table \ref{table:RealResultsGPEKF} and \ref{table:RealResultsGPRTSS} present the mean percentage improvement of the STGP-based trackers over GP-EKF and GP-RTSS, respectively. 
Furthermore, Fig. \ref{fig:RealResults} exhibits the tracking outputs of the algorithms at three different frames of the experiments. 

Contrary to what is shown in \cite[Fig.~11]{aftab2019spatio}, Fig. \ref{fig:RealResults} illustrates that the correct implementation of GP-EKF achieves successful tracking performance for all scenarios. 
In addition, Table \ref{table:RealResultsGPEKF} and \ref{table:RealResultsGPRTSS} indicate that the GP-based trackers outperform their STGP-based counterparts in the accuracy of the kinematic estimates for the scenarios involving the rickshaw and the motorcycle. 
On the other hand, for the pedestrian experiment, the STGP-based trackers are observed to perform better than the GP-based trackers with respect to the kinematic estimates; however, the corresponding levels of improvement is far below than those reported in \cite[Table~V]{aftab2019spatio}. 
Finally, as the precision and recall measures suggest, the shape estimation performance of the algorithms is comparable for all experiments. 

\begin{table}[ht]
    \caption{Mean Percentage Improvements of the STGP-based Trackers over GP-EKF (Real Data)}
    \centering
	\label{table:RealResultsGPEKF}
	\resizebox{\columnwidth}{!}{
    \begin{tabular}{|c|c|l|c|c|c|}
    \hline
    \multicolumn{2}{|c|}{} & \multicolumn{1}{c|}{} & \multicolumn{3}{c|}{\textbf{Scenarios}} \\ \cline{4-6} 
    \multicolumn{2}{|c|}{\multirow{-2}{*}{\textbf{Measure}}} & \multicolumn{1}{c|}{\multirow{-2}{*}{\textbf{Method}}} & \textbf{Rickshaw} & \textbf{Motorcycle} & \textbf{Pedestrian} \\ \hline
     &  & STGP-EKF & \cellcolor[HTML]{FFFFFF}\textbf{-26.66} & \textbf{-21.94} & 32.72 \\ \cline{3-6} 
     & \multirow{-2}{*}{$x$} & \cellcolor[HTML]{C0C0C0}STGP-RTSS & \cellcolor[HTML]{C0C0C0}16.67 & \cellcolor[HTML]{C0C0C0}26.14 & \cellcolor[HTML]{C0C0C0}39.29 \\ \cline{2-6} 
     &  & STGP-EKF & \textbf{-107.59} & \textbf{-22.00} & 18.63 \\ \cline{3-6} 
     & \multirow{-2}{*}{$y$} & \cellcolor[HTML]{C0C0C0}STGP-RTSS & \cellcolor[HTML]{C0C0C0}24.92 & \cellcolor[HTML]{C0C0C0}11.73 & \cellcolor[HTML]{C0C0C0}17.70 \\ \cline{2-6} 
     &  & STGP-EKF & \textbf{-6.81} & \textbf{-37.26} & 30.64 \\ \cline{3-6} 
     & \multirow{-2}{*}{$\dot{x}$} & \cellcolor[HTML]{C0C0C0}STGP-RTSS & \cellcolor[HTML]{C0C0C0}8.82 & \cellcolor[HTML]{C0C0C0}5.44 & \cellcolor[HTML]{C0C0C0}39.41 \\ \cline{2-6} 
     &  & STGP-EKF & \textbf{-31.42} & \textbf{-12.17} & 2.85 \\ \cline{3-6} 
    \multirow{-8}{*}{\rotatebox{90}{RMSE }} & \multirow{-2}{*}{$\dot{y}$} & \cellcolor[HTML]{C0C0C0}STGP-RTSS & \cellcolor[HTML]{C0C0C0}4.35 & \cellcolor[HTML]{C0C0C0}11.94 & \cellcolor[HTML]{C0C0C0}12.38 \\ \hline
    \multicolumn{2}{|c|}{} & STGP-EKF & 0.45 & \textbf{-1.26} & 13.46 \\ \cline{3-6} 
    \multicolumn{2}{|c|}{\multirow{-2}{*}{$P$}} & \cellcolor[HTML]{C0C0C0}STGP-RTSS & \cellcolor[HTML]{C0C0C0}\textbf{-1.07} & \cellcolor[HTML]{C0C0C0}\textbf{-3.59 }& \cellcolor[HTML]{C0C0C0}4.04 \\ \hline
    \multicolumn{2}{|c|}{} & STGP-EKF & \textbf{-7.68} & \textbf{-1.54} & \textbf{-7.32} \\ \cline{3-6} 
    \multicolumn{2}{|c|}{\multirow{-2}{*}{$R$}} & \cellcolor[HTML]{C0C0C0}STGP-RTSS & \cellcolor[HTML]{C0C0C0} 1.99 & \cellcolor[HTML]{C0C0C0}8.35 & \cellcolor[HTML]{C0C0C0}1.03 \\ \hline
    \end{tabular}}
    \end{table}    
    
    \begin{table}[ht]
    \caption{Mean Percentage Improvements of the STGP-based Trackers over GP-RTSS (Real Data)}
    \centering
	\label{table:RealResultsGPRTSS}
	\resizebox{\columnwidth}{!}{
    \begin{tabular}{|c|c|l|c|c|c|}
    \hline
    \multicolumn{2}{|c|}{} & \multicolumn{1}{c|}{} & \multicolumn{3}{c|}{\textbf{Scenarios}} \\ \cline{4-6} 
    \multicolumn{2}{|c|}{\multirow{-2}{*}{\textbf{Measure}}} & \multicolumn{1}{c|}{\multirow{-2}{*}{\textbf{Method}}} & \textbf{Rickshaw} & \textbf{Motorcycle} & \textbf{Pedestrian} \\ \hline
     &  & STGP-EKF & \cellcolor[HTML]{FFFFFF}\textbf{-57.18} & \textbf{-20.75} & 23.03 \\ \cline{3-6} 
     & \multirow{-2}{*}{$x$} & \cellcolor[HTML]{C0C0C0}STGP-RTSS & \cellcolor[HTML]{C0C0C0}\textbf{-3.41} & \cellcolor[HTML]{C0C0C0}26.86 & \cellcolor[HTML]{C0C0C0}30.54 \\ \cline{2-6} 
     &  & STGP-EKF & \textbf{-222.87} & \textbf{-40.25 }& 19.37 \\ \cline{3-6} 
     & \multirow{-2}{*}{$y$} & \cellcolor[HTML]{C0C0C0}STGP-RTSS & \cellcolor[HTML]{C0C0C0}\textbf{-16.77} & \cellcolor[HTML]{C0C0C0}\textbf{-1.48} & \cellcolor[HTML]{C0C0C0}18.45 \\ \cline{2-6} 
     &  & STGP-EKF & \textbf{-16.65} & \textbf{-60.22} & 10.83 \\ \cline{3-6} 
     & \multirow{-2}{*}{$\dot{x}$} & \cellcolor[HTML]{C0C0C0}STGP-RTSS & \cellcolor[HTML]{C0C0C0}0.42 & \cellcolor[HTML]{C0C0C0}\textbf{-10.38} & \cellcolor[HTML]{C0C0C0}22.10 \\ \cline{2-6} 
     &  & STGP-EKF & \textbf{-39.68} & \textbf{-28.90} & \textbf{-12.12} \\ \cline{3-6} 
    \multirow{-8}{*}{\rotatebox{90}{RMSE }} & \multirow{-2}{*}{$\dot{y}$} & \cellcolor[HTML]{C0C0C0}STGP-RTSS & \cellcolor[HTML]{C0C0C0}\textbf{-1.66} & \cellcolor[HTML]{C0C0C0}\textbf{-1.20} & \cellcolor[HTML]{C0C0C0}\textbf{-1.11 }\\ \hline
    \multicolumn{2}{|c|}{} & STGP-EKF & 1.22 & \textbf{-1.07 }& 16.77 \\ \cline{3-6} 
    \multicolumn{2}{|c|}{\multirow{-2}{*}{$P$}} & \cellcolor[HTML]{C0C0C0}STGP-RTSS & \cellcolor[HTML]{C0C0C0}\textbf{-0.31 }& \cellcolor[HTML]{C0C0C0}\textbf{-3.40} & \cellcolor[HTML]{C0C0C0}7.07 \\ \hline
    \multicolumn{2}{|c|}{} & STGP-EKF & \textbf{-10.65} & \textbf{-4.64 }& \textbf{-9.93} \\ \cline{3-6} 
    \multicolumn{2}{|c|}{\multirow{-2}{*}{$R$}} & \cellcolor[HTML]{C0C0C0}STGP-RTSS & \cellcolor[HTML]{C0C0C0}\textbf{-1.29} & \cellcolor[HTML]{C0C0C0}4.93 & \cellcolor[HTML]{C0C0C0}\textbf{-1.81} \\ \hline
    \end{tabular}}
    \end{table}

\section*{Acknowledgment}
We are genuinely grateful to the authors of \cite{aftab2019spatio} for sharing their source code online, which enabled us to prepare this commentary with a reasonable amount of effort. 

\ifCLASSOPTIONcaptionsoff
  \newpage
\fi



%
%
%
\bibliographystyle{IEEEtran}
\bibliography{IEEEabrv,myBiblib}

%







\end{document}